\documentclass[12pt,oneside,english]{article}

 \usepackage{epsfig}
\usepackage[T1]{fontenc}
\usepackage[latin1]{inputenc}
\usepackage{babel}
\usepackage{setspace}

\makeatletter

\begin{document}
 \bigskip  
\centerline{\Large \bf The effects  of  related experiments}
\makeatother

\bigskip \bigskip

\centerline{\large  D.   Bar$^{a}$}  \bigskip \bigskip

\centerline{$^a$Department of Physics, Bar-Ilan University, Ramat-Gan, Israel}

\bigskip \bigskip

\begin{abstract}
{\it  The effects of the experiment  itself   upon the obtained results  and,  
especially, the influence  of a large number of  experiments  
are extensively 
discussed in the literature. 
We show that the important factor that stands at the basis of these effects 
is that the involved experiments are  related and  not 
independent and detached from each other. This relationship takes, as shown 
here,
 different forms for different
situations  and is found in entirely different physical regimes such as  
 the  quantum and classical ones. 
    }
\end{abstract}

\bigskip \bigskip 


\protect \section{Introduction}

The effects of observation upon the obtained results have been extensively
discussed in the literature (see, for example,   \cite{Wheeler} and
references therein). A special kind of experimentations which 
attract many discussions  by many authors  
\cite{Zeno,Simonius,Itano,Aharonov,Facchi} are those  in
which a large number of experiments are involved. Among these one may note the
special role played by those in which the time duration of each of the involved
experiments tends to become infinitesimally small.
Two quantum versions of these very short-time experiments were studied; (1) 
The same
experiment is infinitely repeated, in a finite total time, upon the same system
which results in preserving its initial state (from the very large number of
different states to which it may be projected  by  the experiment) 
  \cite{Zeno,Simonius,Itano}. (2) A very large number of
slightly different experiments are densely performed upon the same system which
results in "realizing"   \cite{Aharonov} the  path of states through which the system
is continuously projected.   That is,   the probability to be  projected 
to this specific path of states (and not to any of the other large number of 
different possible paths  along which the system may evolute) 
  tends to unity   \cite{Aharonov,Facchi}). 
  The first version is termed static Zeno
effect and the second dynamic Zeno effect   \cite{Facchi}.  \par
Another kind of observation that involves many experiments is the  
 space Zeno effect   \cite{Barspace} which    is obtained when one
performs the same experiment in a large number of identical independent
nonoverlapping regions of space. It has been shown   \cite{Barspace} 
that when these regions 
become infinitesimally small, corresponding to the shrinking of the measurement
times in the time Zeno effect, the performance of such experiments has, as for
the static Zeno, a null effect \cite{Barspace}. \par 
We show here that what generally characterizes 
these and other similar  situations is that all the involved experiments, even
those that seems to be  entirely independent, are related to each other 
in some kind of relationship which is responsible for  the obtained results. 
 This is shown for entirely remote and different physical situations which are
studied by different methods such as the Feynman path integral 
\cite{Feynman},   the Everett's universal wave function \cite{Everett,Graham}
and the classical cylinder-piston  system \cite{Reif,Szilard}. 
 We show that  the mentioned  relationship  assumes different forms 
 for these different situations 
 which, actually,    determine  the  necessary details of the involved 
  experiments. Thus, for some situations, like  the static Zeno efect, 
  all the
  systems should be related by preparing them  in the {\it same} initial state 
  whereas   
 for the dynamic Zeno effect   they are related by preparing  them 
 in {\it different} initial states  as we show in
 Section 1.  
  We represent in the following sections examples which
explain the meaning of this relationship and the effects it produces.    
\par 
In Section 2 we use the Feynman path
integral method  \cite{Feynman} to show that if one wants to 
obtain
a large probability for an evolution along a prescribed path of states
then all the involved systems  must be related so that not even 
 two of them happen
to have the same initial state. That is, if this condition is not strictly kept 
and one  prepares these systems so that some of them may  have the same 
initial states then the expected evolution along the specified path of states 
may not
be obtained.  
In Section 3 we use the  Everett's relative state theory  
\cite{Everett,Graham}, which has been especially formulated to take into
account the influence of observers and experiments,  to show  the effect of 
experimenting with  related systems. In Everett's theory the necessity of 
relationship
among the systems is so obvious that it becomes almost trivial to emphasize it. 
We show that if the measurement of the observable $A$ results with 
 $K$ different 
possible outcomes 
  then  the probability to find a specified group
of $r$  eigenvalues (from the given $K$)  in an $n$-sequence  becomes very 
small for large
$K$ and 
small $r$. This is effected through  obtaining an  asymptotically  
large number of different sequences (observers)  for these values of $K$ 
and $r$ which means that the relationship among them is very small.   
   \par  
In Section 4 we   use entropy considerations and  the classical thermodynamical
system of cylinder and  pistons   \cite{Reif,Szilard} to show    
 the influence  of related systems.   
We generalize the discussion in \cite{Szilard} to include a large ensemle of
identical cylinders and show  that the results obtained when these systems
are related greatly differ  from those  obtained when the ensemble's components  
are   independent.  
\protect \section{The Feynman path integrals of the ensemble of observers} 
The large number of experiments discussed here are  
performed by first preparing $N$ similar systems at   
$N$ arbitrarily selected states  from,  actually,  the very large number of
possible states which may be assigned to any system. 
  These systems are then delivered  to the $N$ observers of the ensemble so 
 that  the  system  $i$  $(i=1, 2, \ldots N)$,  prepared at the state $\phi_i$,  
   is assigned to the observer $O_i$. As known   \cite{Merzbacher}, the state of
   any quantum system changes with time without having to touch it.   
   Thus,  we may write for the conditional probability of a self-transition 
    that the
first observer $O_1$ finds his system,   after  
checking its present state, to be at 
the state $\phi_2$ of the second 
observer $O_2$ \begin{equation} \label{e1} \Phi_{12}=\sum_i\phi_{1i}\phi_{i2} 
\end{equation} 
The summation is over all the possible secondary  paths   \cite{Bar0}  
(as those
shown along the middle   path of Figure 1)  
which lead from  $\phi_1$ to  $\phi_2$ and the quantities $\phi_{1i}$ and 
$\phi_{i2}$
denote   \cite{Feynman} the probability amplitudes to proceed from 
the state $\phi_1$ 
to the intermediate one $\phi_i$ and from $\phi_i$ to $\phi_2$ respectively. In the same manner one may
write for the conditional probability amplitude that the second observer $O_2$
finds his system     at the
state $\phi_3$ (of the  observer $O_3$),   provided that
the observer $O_1$ finds  his system   at the state $\phi_2$  
\begin{equation} \label{e2} \Phi_{23|12}=\sum_{ij}\phi_{1i}\phi_{i2}
\phi_{2j}\phi_{j3} 
\end{equation} 
Where $\Phi_{23|12}$ is the remarked conditional probability amplitude and 
$\sum_{ij}$ is the summation over all the secondary  paths  that lead 
from the state $\phi_1$ to $\phi_2$ and over those  from $\phi_2$
to  $\phi_3$. Correspondingly, the conditional probability amplitude that the
$(N-1)$-th observer finds  his system  at the state $\phi_N$ of the observer 
$O_N$  
provided that all the former
$(N-2)$ observers find their respective systems   to be at the   
states $\phi_{i} \ \ (i=2, 3, \ldots N-1)$  is 
\begin{equation} \label{e3} \Phi_{N-1N|12,23,\ldots,N-2N-1}=
\sum_{ij\ldots rs}\phi_{1i}\phi_{i2}\phi_{2j}\phi_{j3}\ldots 
\phi_{N-2r}\phi_{rN-1}\phi_{N-1s}\phi_{sN}, 
\end{equation}  
where the intermediate states in Eqs (\ref{e1})-(\ref{e3}) are orthonormal. 
Figure 1 shows 7  Feynman paths of states, from actually a large number of
paths,  that all
begin at the state $\phi_1$ and end at $\phi_8$  (only 8 states are shown in the figure for
clarity).  The middle path is the specific one
along which the described collective  measurement is performed. 
 Along this line
we have the $N$ ($N=8$  in the figure)  initially prepared states $\phi_1, \phi_2, \ldots \phi_N$ as well
as the secondary Feynman paths that lead from each $\phi_i$ to $\phi_{i+1}$
where $i=1, 2, \ldots,  (N-1)$.  As seen from Eqs (\ref{e1})-(\ref{e3}) 
the  paths (of states)  between
nonneighbouring states as, for example, from $\phi_i$ to $\phi_{i+2}$ are
obtained as  the sum of the separate paths which lead from 
$\phi_i$ to $\phi_{i+1}$ and from $\phi_{i+1}$ to $\phi_{i+2}$. \par  
 The relevant conditional probability is found by
multiplying the last probability amplitude from Eq (\ref{e3})  by its 
conjugate to obtain, omitting
the subscripts of the $\Phi$'s for clarity
\begin{eqnarray} && \Phi^{\dagger}\Phi=\sum_{\grave i \grave j\ldots \grave r 
\grave s}\sum_{ij \ldots rs} \phi_{\grave i1}\phi_{1i}\phi_{2\grave i}\phi_{i2}
\phi_{\grave j2} \phi_{2j}\phi_{3\grave j}\phi_{j3} \ldots 
\phi_{\grave rN-2}\phi_{N-2r}\phi_{N-1\grave r}\phi_{rN-1} \cdot 
\nonumber \\ && \cdot \phi_{\grave sN-1} \phi_{N-1s}\phi_{N\grave s}\phi_{sN} 
=(\sum_{\grave i i}\phi_{\grave i1}\phi_{1i}\phi_{2\grave i}\phi_{i2})
(\sum_{\grave j j}\phi_{\grave j2} \phi_{2j}\phi_{3\grave j}\phi_{j3}) \ldots
 \label{e4} \\ && \ldots (\sum_{\grave r r}\phi_{\grave rN-2}\phi_{N-2r}\phi_{N-1\grave r}\phi_{rN-1})
(\sum_{\grave s s}\phi_{\grave sN-1} \phi_{N-1s}\phi_{N\grave s}\phi_{sN}),  
\nonumber \end{eqnarray}  
where the number of all the double sums $\sum_{\grave i i} \sum_{\grave j j} \ldots
\sum_{\grave r r} \sum_{\grave s s}$ is $N$. \par 
 As remarked, we are interested in
the limit of dense measurement along the relevant Feynman path 
 so we take  $N\to \infty$.  In this limit the length of the secondary
Feynman paths among the initially prepared $N$ states (where  now
 $N \to \infty$) tends to zero   \cite{Bar0} so that  the former 
 probabilities  to proceed along  the secondary paths between the given states 
 become the   
 probabilities  
for these  states \cite{Bar0}.   Thus, 
 we may write for   Eq  (\ref{e4}) in the limit of $N \to \infty$
\begin{eqnarray} && \lim_{N \to \infty}<\!\Phi^{\dagger}|\Phi \!>=
\lim_{N \to \infty}<\! \phi_{\grave I1}| \phi_{2\grave I}\!>
<\!\phi_{I2}|\phi_{1I}\!><\! \phi_{\grave J2}| 
\phi_{3\grave J}\!><\!\phi_{J3}| 
\phi_{2J}\!> \ldots
\nonumber  \\ && \ldots <\! \phi_{\grave R(N-2)}|\phi_{(N-1)\grave R}\!><\!
\phi_{R(N-1)}|\phi_{(N-2)R}\!> <\! \phi_{\grave S(N-1)}| \phi_{N\grave
S}\!>\cdot \label{e5} \\ && \cdot <\!
\phi_{SN}|\phi_{(N-1)S}\!> = 
\delta_{ \phi_{\grave I1}  \phi_{2\grave I}}\delta_{\phi_{1I} \phi_{I2}}
\delta_{ \phi_{\grave J2}  \phi_{3\grave J}}\delta_{\phi_{2J} \phi_{J3}}
\ldots 
\delta_{ \phi_{\grave R(N-2)} \phi_{(N-1)\grave R}} \cdot 
\nonumber \\ && \cdot \delta_{\phi_{R(N-1)} \phi_{(N-2)R}}\delta_{ \phi_{\grave
S(N-1)}
 \phi_{N\grave S}} 
\delta_{\phi_{SN} \phi_{(N-1)S}} =1,   \nonumber 
\end{eqnarray}
where the former indices,  for finite $N$,  $i, {\grave i}, j, {\grave j}, 
\ldots,  r, {\grave r}, s, {\grave s}$ are now, in the limit of $N \to \infty$,
written in an upper case format. This is to emphasize that, unlike the case for
finite $N$,   neighbouring states along the traversed Feynman path 
differ infinitesimally.   
The last result of unity follows because, as just noted,  in the limit of $N \to \infty$
successive states differ infinitesimally from each other so we may write  
as in   \cite{Facchi}  
$<\! \phi_{\grave k-1}| \phi_{\grave k}\!>=<\!\phi_{k-1}|\phi_{k}\!> 
 \approx \delta_{ \phi_{\grave k-1} \phi_{\grave k}}=
\delta_{\phi_{k-1}\phi_k}\approx 1$.  Thus, we see that performing dense 
measurement along any Feynman path of states 
results
in its "realization"  \cite{Aharonov,Facchi}  in the sense that the probability to proceed through 
 all of its  states
 tends to unity.\par
As remarked, the key feature of the described dense measurement is that all the
$N$ systems  are related to each other in such a way that 
 their $N$ initial states are  prepared to
be different from each other where in the limit of $N \to \infty$ these
differences become infinitesimal for neighbouring states. Note that we do not
take all the $N$ initial states to be identical since in this case all the
former discussion and Eqs (\ref{e1})-(\ref{e5}) would not be relevant. This is
because the primary Feynman path formerly applied  for describing the path of these $N$
states would shrink to a point if they are identical. Note that by taking the
limit of $N \to \infty$ and by having (for continuity)  a slight differences 
between neighbouring states we have already caused the secondary Feynman paths
of the relevant primary one (see Figure 1) to shrink and disappear. Thus, as
noted, taking $N$ identical initial states may causes the primary Feynman path,
in the limit of $N \to \infty$,  to shrink to a point which is not the meant
results of this discussion. Note that this procedure of taking $N$ different
initial states where the neighbouring ones  differ infinitesimally in the
limit of $N \to \infty$ is the key property of the dynamic Zeno effect as seen
in \cite{Aharonov,Facchi} (see, especially, Sections 1-2 in \cite{Facchi}). Also
the continuity condition is not violated as seen from Eq (\ref{e5}). 
\par  
  Note that the described dense measurement  is
performed through the joint action of all  the members of the ensemble     
   as  schematically illustrated in  Figure 2 which  represents 
     the 
ensemble of observers  after the remarked collective  measurement. 
Each 
batch of 4 similar curves  denotes  a member of the 
$N$-ensemble system  that  has, as known,  a large number of 
different possible Feynman paths of evolution  
(only 4 are shown for clarity).  In the 
middle 
part of the figure 
we have a large number of different batches of paths all mixed  among them  
 so 
it becomes  difficult to see  which curve belongs to which batch. This
corresponds to  densely measuring  ( $N \to \infty$)   where neighbouring states 
infinitesimally differ from each other. 
The emphasized
path in Figure 2 is the definite  path along which the described
collective dense  measurement has been done. Note that this path, actually,
belongs to all the different mixed  batches which means that  
after completing  the  collective 
measurement each one of those that participates in it has now the same 
Feynman path.  The reason is that although each  observer $O_i$ of the ensemble 
performs his  experiment on   his
 prepared state $\phi_i$, nevertheless, the results he obtains are
valid for all the others  since any  observer 
that
acts on the same state $\phi_i$ under exactly the same conditions obtains 
 the same result.   
  In other words,   
the emphasized  Feynman path belongs now to  all of them in 
the
sense that the probability for each to move  along its 
 constituent states tends to unity   
as seen from Eq (\ref{e5}).  \par 

We note that in contrast to the relationship discussed just now 
which demands a preparation of different initial states for the 
realization of its (dynamic Zeno)  effect the situation regarding the 
static Zeno effect is opposite and 
contrary. This is because the required relationship there demands to prepare all
the initial states of the involved experiments to be identical \cite{Zeno} so 
as to
be able to preserve this state in time. We note that 
 using a large ensemble of similar systems for analysing 
experimental results  
has been fruitfully done in the 
literature  \cite{Graham,Finkelstein,Hartle,Smolin} without invoking any Zeno
idea.  
It has been shown, for example, that by considering  $N$ {\it identical} 
systems all prepared 
in the same initial state one may derive the probability interpretation of quantum 
mechanics in the limit of $N \to \infty$. That is, this probability  
is not imposed 
upon the theory as an external assumption as done in the conventional Copenhagen 
interpretation \cite{Merzbacher} of quantum mechanics  but is derived from other principles of quantum 
mechanics   \cite{Smolin}. This is done using Finkelstein theorem  
\cite{Finkelstein,Smolin}.

\begin{figure}[hb]
\begin{minipage}{.65\linewidth}
\centering\epsfig{figure=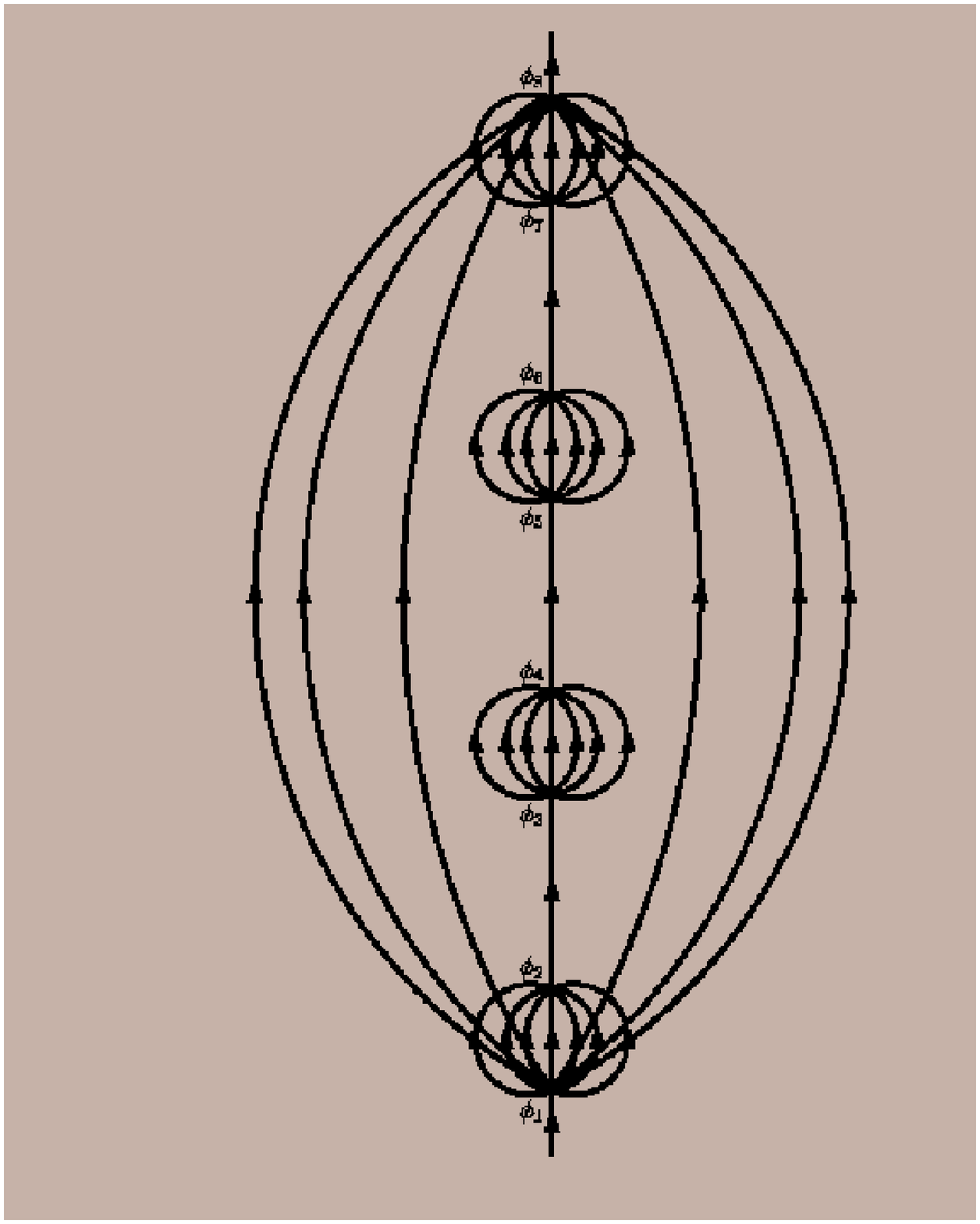,width=\linewidth}
\caption{Seven  Feynman paths of states, from a very large number of
possible ones,  
that all begin at the state $\phi_1$ (at the bottom) and end
 at $\phi_8$ are shown in the figure (only 8 states are shown for clarity).  
 The middle path is the one along 
which   the collective dense measurement is performed by the ensemble
members $O_i, \ \ \  i=1, 2, \ldots N$.   The $N$  separate systems of these 
observers have been initially prepared in the
states $\phi_i \ \ \ i=1, 2, \ldots N$.  
 Note the secondary Feynman paths between neighbouring states in the
middle path.} 
\end{minipage}
\end{figure}

\begin{figure}[hb]
\begin{minipage}{.70\linewidth}
\centering\epsfig{figure=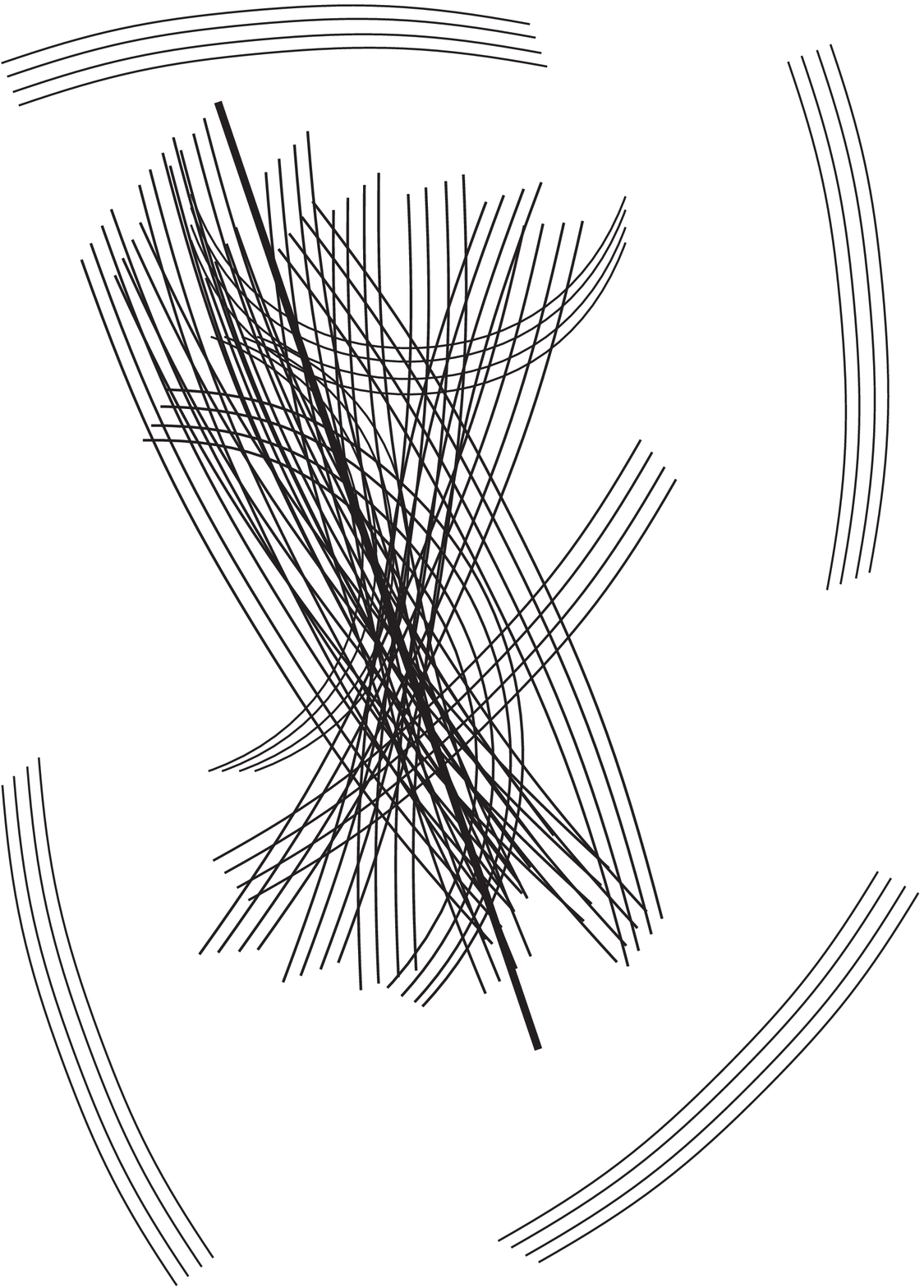,width=\linewidth}

\caption{A schematic representation of the physical situation after
performing  the collective 
dense measurement symbolized by   Figure 1. Note that  although no member of 
the 
ensemble has done dense
measurement by himself, nevertheless, the joint action of all or most of the
observers has resulted in ``realizing'' the specific Feynman path from Figure
1
for {\it all the participating observers}. This "realized"  path is shown emphasized
in the figure.}
\end{minipage}
\end{figure}

\protect \section{\label{sec1}  The relative state theory of Everett}
  The last results may be demonstrated in a more natural and
appealing manner by using the relative state theory of Everett 
  \cite{Everett,Graham} which has been
formulated, especially, for taking observers into account. We use, in the
following, the special notation and terminology of this theory. Thus, if the
initial state was some eigenstate of an operator $A$ the total initial state of
the (system $S$ $+$ observer $O$) is denoted by $\Psi_i^{S+O}=\phi_i\Psi^O[...]$,
where $\phi_i$ is the initial eigenstate of the system $S$ and $\Psi^O[...]$ 
denotes the observer's state before the measurement. After the experiment the
observer's state is denoted by $\Psi^O[...\alpha_i]$, where $\alpha_i$ stands
for recording  the eigenvalue $\alpha_i$ by the observer and   the total final
state of  the (system $S$ $+$ observer $O$) 
is $\Psi_f^{S+O}=\phi_i\Psi^O[...\alpha_i]$.  Now, if the initial state of 
the system is not an
eigenstate then it may be expressed as a superpositions of such eigenfunctions 
  $\sum_ia_i\phi_i$  and the total states
before and after the measurement are   \cite{Everett,Graham}  
$\Psi_i^{S+O}=\sum_ia_i\phi_i\Psi^O[...]$,  and 
$\Psi_f^{S+O}=\sum_ia_i\phi_i\Psi^O[...\alpha_i]$ 
 respectively where $a_i=<\!\phi_i|\Psi^{S+O}\!>$. 
 We note that we consider here the one-step measurement of \cite{Everett} 
 and not the two-step version \cite{Graham} of it in which a macroscopic apparatus is
 introduced between a microscopic system and a macroscopic observer. \par
 We, now, wish to represent the former process of  measuring 
   the observable $A$
 on $N$ identical independent systems.
 We assume that the initial state of each one of the $N$ systems is not an
 eigenstate of $A$ so it can be expanded as a superpositions of such
 eigenfunctions. Thus, we may write for the initial state of the $N$-system
 ensemble  
  \cite{Everett}
 \begin{equation} \label{e6} 
 \Psi_i^S=\sum_i\sum_j\cdots  \sum_k\sum_l
 <\!\phi_i|\phi\!><\!\phi_j|\phi\!> \ldots
 <\!\phi_k|\phi\!><\!\phi_l|\phi\!> \phi_i\phi_j \ldots \phi_k \phi_l
   \end{equation}
 where $\phi_i, \ \ \phi_j, \ \ldots \phi_k, \ \ \phi_l$  are the 
 eigenfunctions of the operator $A$. Thus, the initial and final states of the
 total system ($N$ systems $+$  observer)  are
  \begin{equation} \label{e7} 
 \Psi_i^{S+O}= \sum_i\sum_j\cdots  \sum_k\sum_l
 <\!\phi_i|\phi\!><\!\phi_j|\phi\!> \ldots
 <\!\phi_k|\phi\!><\!\phi_l|\phi\!> \phi_i\phi_j \ldots \phi_k
 \phi_l\Psi^O[\dots]
   \end{equation} 
   \begin{eqnarray} && 
 \Psi_f^{S+
 O}=\sum_i\sum_j\cdots  \sum_k\sum_l
 <\!\phi_i|\phi\!><\!\phi_j|\phi\!> \ldots
 <\!\phi_k|\phi\!><\!\phi_l|\phi\!> \phi_i\phi_j \ldots    \label{e8} \\
 && \ldots \phi_k
 \phi_l\Psi^O[\alpha_i,\alpha_j\ldots \alpha_k,\alpha_l] \nonumber 
   \end{eqnarray} 
 where $\Psi^O[\alpha_i,\alpha_j \ldots \alpha_k,\alpha_l]$ denotes that  the
 observer has measured the $n$ eigenvalues  $ \alpha_i,\alpha_j \ldots
 \alpha_k, \alpha_l$   
   of  $A$. Note that each term in Eq (\ref{e8}) actually    denotes an observer
 with his specific  sequence  $[\alpha_i,\alpha_j\ldots \alpha_k,\alpha_l]$ 
 which  results 
 from the $n$ experiments. Thus, Eq (\ref{e8}), termed the Everett's universal 
 wave function   \cite{Everett,Graham}, gives all the possible results 
 that may be obtained from
 performing  the same experiment upon the $n$ systems. \par   
  We, now,  count the number of observers
 that have the same or similar  sequences  
 $[\alpha_i,\alpha_j\ldots \alpha_k,\alpha_l]$ which  record, as remarked, 
 the $n$
 measured eigenvalues. For this we assume that each measurement of  the
 observable $A$  may give  any of  $K$ possible different eigenvalues and that 
 some of the $n$ components in any sequence  may be identical.  
 Thus,
 denoting by  $R_1,R_2,\ldots,R_r$ the numbers of times 
 the $r$ particular different eigenvalues
 $l_1,l_2,\ldots,l_r$  respectively appear  in some specified sequence 
 $[\alpha_i,\alpha_j\ldots \alpha_k,\alpha_l]$ one may see from Eq (\ref{e8}) that
 each possible value of $R_i$ in the range $0 \le R_i \le n$, and for each 
 $i
 \ \ \ 
 (1 \le  i \le r)$, may be realized in some observer.  
   Now, the number of
 sequences in which $l_1,l_2,\ldots,l_r$  respectively  occur  at 
 $R_1,R_2,\ldots,R_r$ predetermined positions is 
 $(K-r)^{(n-\sum_{i=1}^{i=r}R_i)}$.   
 This is because  for each position in the sequence 
  $[\alpha_i,\alpha_j\ldots \alpha_k,\alpha_l]$ in which  
   the $r$ 
  eigenvalues 
  $l_1,l_2,\ldots,l_r$ are  
  absent there are $(K-r)$ possible locations. Note that $K$ and $r$ should
  satisfy the relation $K \ge r$.   
  Thus,  the total number of sequences 
  in which 
$l_1,l_2,\ldots,l_r$    respectively  occur  in  $R_1,R_2,\ldots,R_r$ 
positions (we
denote this number by $N_{l_1,l_2,\ldots,l_r}$) is 
\begin{eqnarray} &&  N_{l_1,l_2,\ldots,l_r}=\left( \begin{array}{c} n \\ R_1 
\end{array} \right)\left( \begin{array}{c} (n-R_1) \\ R_2 
\end{array} \right)\left( \begin{array}{c}(n-(R_1+R_2)) \\ R_3 
\end{array} \right)\ldots \label{e9} \\ && \ldots \left( \begin{array}{c} (n-\sum_{i=1}^{i=r-1}R_i) \\ 
R_r \end{array} \right)(K-r)^{(n-\sum_{i=1}^{i=r}R_i)}, \  \ K \ne r, \ \ 
K \ne 1 
\nonumber  
\end{eqnarray} 
where $\left( \begin{array}{c} n \\ R_1 
\end{array} \right)$ is the number of possible ways to choose in the $n$-member
sequence $[\alpha_i,\alpha_j\ldots \alpha_k,\alpha_l]$ $R_1$ places for $l_1$, 
$\left( \begin{array}{c} (n-R_1) \\ R_2 
\end{array} \right)$ is the number of possible ways to choose $R_2$ places from
the remaining $(n-R_1)$ etc. Note that when  $K=r$, 
which means that any one of the $K$ possible results of  the experiment must
be one   of the $r$ eigenvalues $l_1,l_2,\ldots,l_r$, then the probability 
that  all the $n$ components (where some of them may be identical)  
 of  any sequence  belong to the 
$l_1,l_2,\ldots,l_r$'s group  is
unity.  In this case  the number of observers  that have in their 
sequences all the $r$ different  eigenvalues $l_1,l_2,\ldots,l_r$ is 
\begin{eqnarray*} &&   N_{l_1,l_2,\ldots,l_r}=\left( \begin{array}{c} n \\ R_1 
\end{array} \right)\left( \begin{array}{c} (n-R_1) \\ R_2 
\end{array} \right)\left( \begin{array}{c}(n-(R_1+R_2)) \\ R_3 
\end{array} \right) \ldots \left( \begin{array}{c} 
(n-\sum_{i=1}^{i=r-1}R_i) \\ 
R_r \end{array} \right), \nonumber \\ &&  K = r, \ \ \ K \ne 1 
 \end{eqnarray*} which is  the same as  Eq
(\ref{e9}) but without the factor in $K$. \par 
  The
relevant measure may be found   \cite{Graham} by taking account 
of the expected
relative frequency $P_{l_1,l_2,\ldots,l_r}$  of the 
eigenvalues $l_1,l_2,\ldots,l_r$  and the
corresponding relative frequency $Q_{m \ne l_1,l_2,\ldots,l_r}$ 
of any other eigenvalue $m$ different from 
$l_1,l_2,\ldots,l_r$.   The first is given by    
$P_{l_1,l_2,\ldots,l_r}=|\!\!<\!\Psi_{l_1,l_2,\ldots,l_r}|\Psi\!>\!\!|^2$ 
 where
 $|\Psi_{l_1,l_2,\ldots,l_r}\!>$ is the state in which the eigenvalues $l_1,
 l_2, \ldots, l_r$ occur among those of the sequence  
 $[\alpha_i,\alpha_j\ldots \alpha_k,\alpha_l]$ and the second is   
$Q_{m\ne l_1,l_2,\ldots,l_r} =\sum_{(m\ne l_1,l_2,\ldots,l_r)}|\!\!<\!\Psi_{m \ne
l_1,l_2,\ldots,l_r}|\Psi\!>\!\!|^2=1-
P_{l_1,l_2,\ldots,l_r}$ were $|\Psi_{m \ne l_1,l_2,\ldots,l_r}\!>$  
is the state in which the eigenvalues $l_1,
 l_2, \ldots, l_r$  do not occur among those of this sequence. Thus,  
 the measure of the  sequences that have 
the eigenvalues $l_1,l_2,\ldots,l_r$  at the respective $R_1,R_2,\ldots,R_r$ predetermined
positions       
 is $P_{l_1,l_2,\ldots,l_r}^{\sum_{i=1}^{i=r}R_i}Q_{m \ne l_1,l_2,\ldots,l_r}^{(n-\sum_{i=1}^{i=r}R_i)}$. 
  The last expression 
 must be multiplied by the number of  possible ways to choose first $R_1$
 places for $l_1$ from the  $n$ positions of the sequence  
 $[\alpha_i,\alpha_j\ldots \alpha_k,\alpha_l]$, then to choose $R_2$ places for $l_2$
 from the
 remaining $n-R_1$ etc,  
  until the last step of choosing $R_r$ places from   
 $(n-\sum_{i=1}^{i=r-1}R_i)$  (see  Eq (\ref{e9})).   
 That is, the sought-for measure $M_e$ is  
\begin{eqnarray} && M_e(r)=\left( \begin{array}{c} n \\ R_1 
\end{array} \right)\left( \begin{array}{c} (n-R_1) \\ R_2 
\end{array} \right)\left( \begin{array}{c} (n-(R_1+R_2)) \\ R_3 
\end{array} \right)\ldots \label{e10} \\ && \ldots \left( \begin{array}{c} (n-\sum_{i=1}^{i=r-1}R_i) \\ 
R_r \end{array} \right)
P_{l_1,l_2,\ldots,l_r}^{\sum_{i=1}^{i=r}R_i}Q_{m \ne l_1,l_2,\ldots,l_r}^{(n-\sum_{i=1}^{i=r}R_i)}, 
\nonumber  
\end{eqnarray} which is the Bernoulli distribution \cite{Spiegel}. As remarked
in \cite{Graham} $M_e(r)$ from Eq (\ref{e10}) may be approximated, for large
$N$, by a Gaussian distribution with mean $NP_{l_1,l_2,\ldots,l_r}$ and standard
deviation $\sqrt{NP_{l_1,l_2,\ldots,l_r}Q_{m \ne l_1,l_2,\ldots,l_r}}$.  
  We, now,  calculate  
an explicit expression  for $P_{l_1,l_2,\ldots,l_r}(r)$ and 
$Q_{m \ne l_1,l_2,\ldots,l_r}(r)$ as functions of
$r$, for  
$n=30$.   The probability $P_{l_1,l_2,\ldots,l_r}(r)$ to find
the eigenvalues   
$l_1,l_2,\ldots,l_r$ among those  of the sequence 
$[\alpha_i,\alpha_j\ldots \alpha_k,\alpha_l]$  may be written as 
$P_{l_1,l_2,\ldots,l_r}(r)=|<\!\Psi_{l_1,l_2,\ldots,l_r}|\Psi\!>|^2=\frac{r}{n}=
\frac{r}{30}$,  and the probability to find any other eigenvalue 
$m \ne l_1,l_2,\ldots,l_r$  is 
$Q_{m \ne l_1,l_2,\ldots,l_r}(r)=\sum_{(m\ne l_1,l_2,\ldots,l_r)}|<\!\Psi_{m \ne
l_1,l_2,\ldots,l_r}|\Psi\!>|^2=1-
P_{l_1,l_2,\ldots,l_r}=1-\frac{r}{30}=\frac{(30-r)}{30}$. 
 For 
simplifying 
the following calculations we assign to all the  values of 
$R_i \  \ i=1,2,\ldots r$ the unity value, in which case each of the given
eigenvalues $l_1,l_2,\ldots,l_r$ may occurs only once in the sequence 
$[\alpha_i,\alpha_j\ldots \alpha_k,\alpha_l]$.   
   Thus,  the  relevant total number of sequences (observers) 
 $N_{l_1,l_2,\ldots,l_r}(K,r)$  and the corresponding measure $M_e(r)$  
 from Eqs (\ref{e9})-(\ref{e10}) 
are given by  
 \begin{eqnarray} &&  N_{l_1,l_2,\ldots,l_r}(K,r)=\left(
 \begin{array}{c} 30 
 \\ 1 
\end{array} \right)\left( \begin{array}{c} 29 \\ 1 
\end{array} \right)\ldots \left( \begin{array}{c} (30-(r-1)) \\ 1 
\end{array} \right) \cdot \label{e11} \\ && \cdot (K-1)^{(30-r)}= 
\prod_{i=0}^{i=r-1}(30-i)(K-1)^{(30-r)}
\nonumber \end{eqnarray}   and 
  \begin{eqnarray} && 
M_e(r)=\left( \begin{array}{c} 30 
 \\ 1 
\end{array} \right)\left( \begin{array}{c} 29 \\ 1 
\end{array} \right)\ldots \left( \begin{array}{c} (30-(r-1)) \\ 1 
\end{array} \right)(\frac{r}{30})^r  \cdot \label{e12}  \\ && \cdot 
(\frac{(30-r)}{30})^{(30-r)}
= \prod_{i=0}^{i=r-1}(30-i)
(\frac{r}{30})^r(\frac{(30-r)}{30})^{(30-r)} \nonumber 
\end{eqnarray} 

 \begin{table}
\caption{\label{table1} The table shows the  number of observers that have $r$
    positions in their 30 places sequences occupied by the preassigned eigenvalues, where
    the numbers $K$ of possible values for each experiment are 1100, 100, 10, 5
    and 2. The untabulated places for $K=10$, $K=5$ and $K=2$ are when $K \le
    r$.}
     \begin{center}
      \begin{tabular}{c|c|c|c|c|c|} 
        \  r&Number \ \ of  &Number \ \ of  &Number \ \ of &
       Number \ \ of  &Number \ \ of  \\ &observers for&
      observers for& 
      observers for & observers for&observers for \\&\ \ K=1100&\ \ K=100&\ \ 
      K=10&\ \ K=5&\ \ K=2\\
   \hline \hline
 
$  1  $&$      4.6350491\cdot 10^{89}    $&$    2.2415163\cdot 10^{59}$&$1.4130386\cdot 10^{29} $&$ 8.6469113\cdot  10^{18}$&$3 \cdot 10^{1}$\\
$  2  $&$      1.1922979\cdot 10^{88}    $&$    4.9413927\cdot 10^{58}$&$1.6828247\cdot 10^{28}$&$ 1.9902809\cdot 10^{16}$&$$870\\
$  3  $&$     2.9665811\cdot 10^{86}     $&$   1.0703212\cdot 10^{58}$&$1.6007531\cdot 10^{27} $&$  3.2695439\cdot 10^{12}$&$---$\\
$  4   $&$      7.1304257\cdot 10^{84}   $&$    2.2755852\cdot 10^{57}$&$1.1219501\cdot 10^{26} $&$  6.5772\cdot 10^{5}$&$---$\\
$  5    $&$     1.6533598\cdot 10^{83}    $&$   4.7435614\cdot10^{56}$&$5.0964117\cdot 10^{24} $&$ 171\cdot 10^{5}$&$---$\\
$  6    $&$     3.6929220\cdot 10^{81}    $&$   9.6832889\cdot 10^{55}$&$1.2033562\cdot 10^{23} $&$ ---$&$---$\\
 $ 7      $&$   7.9328524\cdot 10^{79}     $&$  1.9331852\cdot 10^{55}$&$ 9.6594968\cdot 10^{20} $&$ ---$&$---$\\
$  8     $&$    1.6360310\cdot 10^{78}   $&$     3.7689961\cdot 10^{54}$&$9.8981353\cdot 10^{17} $&$ ---$&$---$\\
$  9    $&$     3.2332250\cdot 10^{76}   $&$     7.1644920\cdot 10^{53}$&$5.1917786\cdot 10^{12} $&$ ---$&$---$\\
 $ 10   $&$    6.1103406\cdot 10^{74}   $&$     1.3255181\cdot 10^{53}$&$1.09027\cdot 10^{14}$&$---$&$---$\\
 $ 11    $&$   1.1017807\cdot 10^{73}    $&$      2.3821855\cdot 10^{52}$&$---$&$---$&$---$\\
 $ 12    $&$   1.890772\cdot 10^{71}     $&$    4.1496088\cdot 10^{51}$&$---$&$---$&$---$\\
 $ 13     $&$   3.0795964\cdot 10^{69}   $&$     6.9890614\cdot 10^{50}$&$---$&$---$&$---$\\
 $ 14    $&$     4.7458902\cdot 10^{67}  $&$     1.1350519\cdot 10^{50}$&$---$&$---$&$---$\\
 $ 15     $&$    6.8961474\cdot 10^{65}  $&$    1.771921\cdot 10^{49}$&$---$&$---$&$---$\\
$  16    $&$   9.4115613\cdot 10^{63}   $&$    2.6494866\cdot 10^{48}$&$---$&$---$&$---$\\
 $ 17    $&$   1.2010185\cdot 10^{62}   $&$  3.7791572\cdot 10^{47}$&$---$&$---$&$---$\\
 $ 18    $&$   1.4257725\cdot 10^{60}  $&$    5.1178711\cdot 10^{46}$&$---$&$---$&$---$\\
 $ 19    $&$    1.5652618\cdot 10^{58} $&$     6.5439439\cdot 10^{45}$&$---$&$---$&$---$\\
 $ 20   $&$    1.5781003\cdot 10^{56}  $&$    7.8486852\cdot 10^{44}$&$---$&$---$&$---$\\
$  21   $&$    1.4490723\cdot 10^{54}  $&$    8.7607415\cdot 10^{43}$&$---$&$---$&$---$\\
$  22   $&$    1.1997469\cdot 10^{52}  $&$    9.0135604\cdot 10^{42}$&$---$&$---$&$---$\\
$  23   $&$     8.8458478\cdot 10^{49}  $&$    8.4462633\cdot 10^{41}$&$---$&$---$&$---$\\
$  24  $&$      5.7174346\cdot 10^{47}  $&$    7.0991954\cdot 10^{40}$&$---$&$---$&$---$\\
$  25   $&$     3.1733732\cdot 10^{45}  $&$    5.2454789\cdot 10^{39}$&$---$&$---$&$---$\\
$  26  $&$        1.4705031\cdot 10^{43} $&$    3.3141771\cdot 10^{38}$&$---$&$---$&$---$\\
$  27   $&$      5.4614504\cdot 10^{40} $&$    1.7197979\cdot 10^{37}$&$---$&$---$&$---$\\
$  28   $&$     1.5241217\cdot 10^{38}  $&$    6.8753541\cdot 10^{35}$&$---$&$---$&$---$\\
 $ 29   $&$     2.8408581\cdot 10^{35}  $&$    1.8832953\cdot 10^{34}$&$---$&$---$&$---$\\
 $ 30  $&$       2.6525286\cdot 10^{32} $&$     2.6525286\cdot 10^{32}$&$---$&$---$&$---$\\

\end{tabular} 
\end{center}
\end{table}

 In Table 1 we show the number of observers
(sequences) that have $r$
predetermined different eigenvalues in their respective $n$-place  sequences 
 for $n=30$ and for the      five  different values 
of $K=1100, 100, 10, 5, 2$.        Note that for the large values of 
$K$, which
signifies  a large  number of  possible results for the measurement of the
observable $A$,   the sequences most frequently encountered are, as expected,  
the 
 ones that contain 
 small number of the $r$ eigenvalues.       That is, the larger is $K$ 
the smaller is the relationship among the ensemble's 
members. For example, for $K=1100$ and  $K=100$ the
number of different observers (sequences) with $r=1$,  that have only  
 one of the preassigned 
 eigenvalues, are $4.6350491 \cdot 10^{89}$ and  
$2.2415163 \cdot 10^{59}$  respectively  compared  to 
$2.6525286 \cdot 10^{32}$ and  
$2.6525286 \cdot 10^{32}$ that have all  the 30 places in their 
sequences   occupied by
such eigenvalues.  That is, for $K=1100$ and  $K=100$ the number of different observers 
(sequences) with $r=1$ 
are respectively larger  by the  factors  of $1.7474 \cdot 10^{57}$ and 
$8.45048 \cdot 10^{26}$  compared to  those with $r=30$.  \par 
These  results, although in a smaller scale, 
 are found  also for small  $K$ which 
signifies a small number of possible different  results 
for the measurement of $A$.    That is, most  observers  
are found to have in their sequences  a small 
 number of the $r$ predetermined eigenvalues. Note that for small $K$ we can
 read from Table 1  the values of  $N_{l_1,l_2,\ldots,l_r}$ also for  $K=r$.   
 For example, for 
 $K=r=2$ the number of different sequences is greater by a factor of 29 than for
 $K=2$ and $r=1$. 
The results of  Table 1 are
corroborated by directly calculating  the
relative rate $R(K,r)$ of the increase of $N_{l_1,l_2,\ldots,l_r}$  from Eq
(\ref{e11}) which  is 
\begin{equation} \label{e13} R(K,r)= \frac{N_{l_1,l_2,\ldots,l_r}(K,r)-
N_{l_1,l_2,\ldots,l_r}(K,r-1)}{N_{l_1,l_2,\ldots,l_r}(K,r)},  \end{equation} 
It has been found that the rate  $R(K,r)$  is always negative for the order of
magnitudes of $K=100$ and $r \le K$ discussed here which means that 
 $N_{l_1,l_2,\ldots,l_r}(K,r) < N_{l_1,l_2,\ldots,l_r}(K,r-1)$. That is, 
 as we have found from Table 1, the large number of observers (sequences) are
 found at small $r$. Also, we find for small $r$ (not shown) that 
 the larger $K$ becomes  the more inclined toward negative values is 
 the  surface of   $R(K,r)$ which means that the large number of observers are
 found, as in Table 1,  at large $K$ and small $r$.   
  When $K=1$,  which means that
there is only one result for the measurement of $A$,  then 
 we must have $r=1$ and the former problem
of calculating the probability to find $r$ specified eigenvalues in $n$-sequence
reduces to finding one known eigenvalue  which is trivially unity since there
exists no other eigenvalue to measure.   
\par 
In summary,  we see that an important necessary aspect for obtaining a large
probability for a specific configuration of $n$-sequence is that its components
must be   {\it related}. This relationship is expressed through the number of
different sequences in Table 1 so that the smaller is this number the greater is
the relationship and vice versa. The number of different sequences (observers) 
  is determined  by  $K$ and $r$ so 
 that for  small $K$ and large $r$,  where always $K \ge r$,  this number is small 
 and for large $K$ and small $r$ it is large.     
 Note that if they do not  measure the same observable 
then the observers are  totally unrelated and our former results would not be 
obtained 
  even for small $K$.    
 \bigskip
 \section{\label{sec9}The classical effect of an ensemble of 
  observers}   
  We discuss now the same system used in  \cite{Szilard} for demonstrating the effect 
of observation upon the experimental results.  The discussion 
in  \cite{Szilard} is generalized to include 
the large ensemble of related  $N$ thermodynamical systems,  of the
 kind studied  in   \cite{Szilard}.  
  That is,    a hollow cylinder that contains  $n$
  particles, not all of the same kind,  among    
     four pistons as
 shown in Figure 3.  The pistons $A$ and $\grave A$ are fixed while $B$ and
 $\grave B$ may move along the cylinder. Also  the pistons $\grave A$ and 
 $B$  do not allow passage of particles through them, whereas $A$ and $\grave
 B$ are permeable so that each permits some kind of particles to move
 through it where those that are allowed to pass through $A$ are not allowed
 through $\grave B$ and vice versa. The pistons $B$ and $\grave B$ move in such
 a way that the distances $B\grave B$ and $A\grave A$ are always equal as seen
 in Figure 3. These
 distances are measured using the $x$ axis which is assumed to be upward along
 the cylinder. 
 We assume that the piston $A$ is permeable only to  the particles inside the
 interval $(x_1,x_2)$ and $\grave B$  only to those outside it.  We
 denote by $w_1$ the initial probability  
  that any randomly selected particle is found
 to be in the interval $(x_1,x_2)$ and by $w_2$ that it is outside it. 
At  first the pistons $B$ and $\grave B$ were at the positions of 
$A$ and $\grave A$ respectively
 and all the $n$ particles were in the one space between. 
 We, now, wish to  perform, reversibly and with no
 external force, a complete cycle of first moving up the pistons $B \grave B$
 and then retracing them back to their initial places.   
 Thus,  by  moving up, without
 doing work, the pistons $B$ and $\grave B$  the volume
 enclosed between them equals, as remarked,  that between $A\grave A$ 
and  we obtain two separate equal volumes, each of which equals to
 the initial one.     Now,  since $A$ is
 permeable to the particles in the interval $(x_1,x_2)$ and $\grave B$ to the
 rest the result is that the upper volume  $B\grave B$ contains only the particles from
 the predetermined interval $(x_1,x_2)$ and the lower $A\grave A$ only the
 others.  \par

 \begin{figure}[ht]
\begin{minipage}{.50\linewidth}
\centering\epsfig{figure=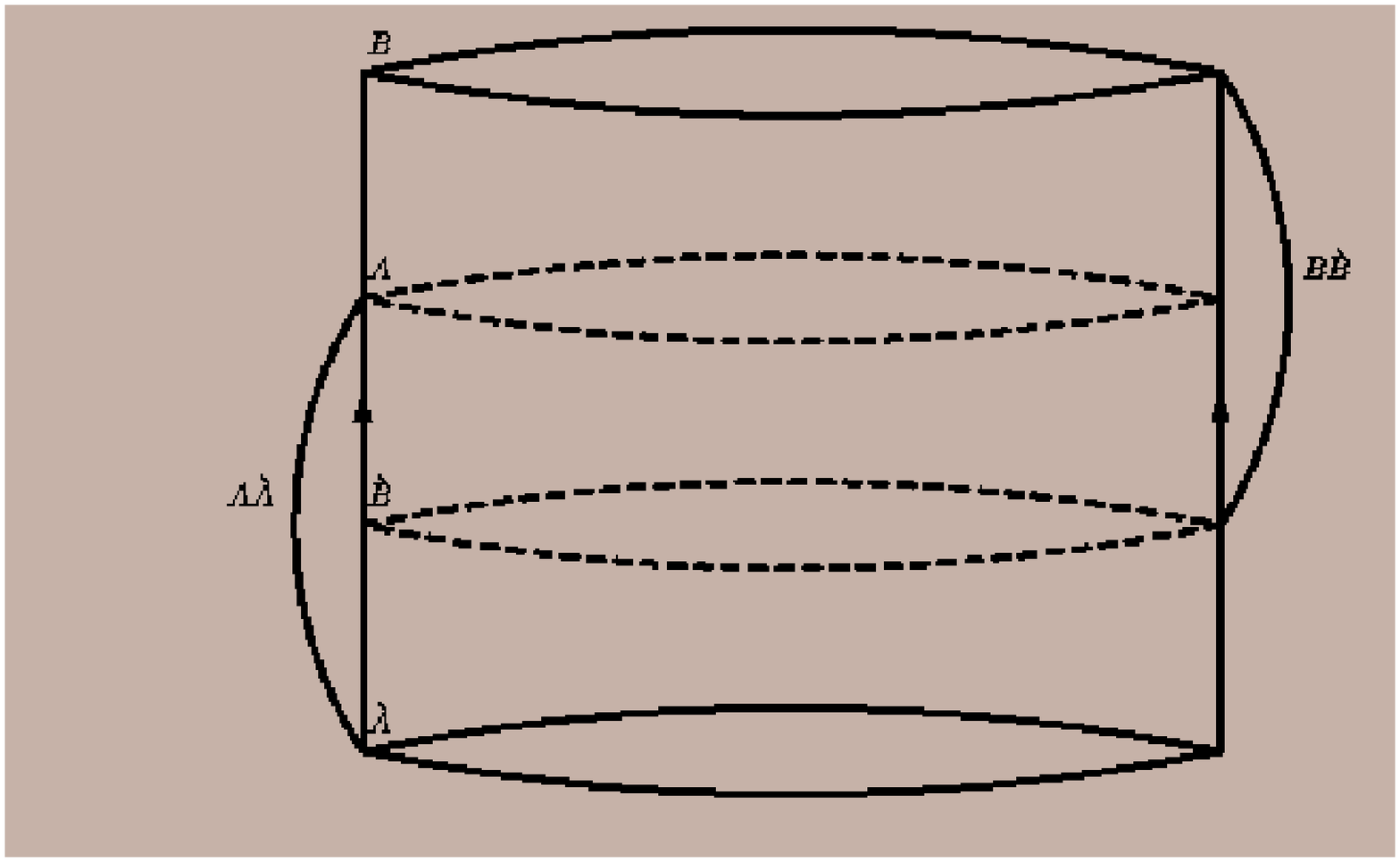,width=\linewidth}

\caption{The cylinder with the four pistons. The pistons $A$ and 
$\grave A$ are
fixed  whereas $B$ and $\grave B$ may move along the cylinder.  Also the piston $A$
is permeable to the molecules inside the interval $(x_1,x_2)$ (see text) and $\grave B$ to
those outside it.  } 
\end{minipage}
\end{figure}

 When we retrace the former steps and   move down 
   the pistons $B$ and $\grave B$ to their former  places at  $A$ and $\grave A$
  the same initial volume is obtained.  We must  take into account, however,   that  
  during the upward motion  some particles that were 
  inside (outside) 
 the interval $(x_1,x_2)$ may come out of (into) it 
  due to thermal or other kind of
 fluctuation so that  these particles change 
  from the kind that may pass through the piston $A$  ($\grave B$)  
  into the kind
 that is not allowed to do that. Thus, the last step of retracing the pistons
 $B$, $\grave B$ into their former initial positions at the pistons $A$, $\grave
 A$ respectively can not be performed without doing work since the molecules
 that have come out of (into) the  interval $(x_1,x_2)$ are not permitted now
 to pass through $A$ ($\grave B$). That is,  the former process of expanding 
 the volume is not reversible as
 described because we have to exert force on these molecules to move them back
 into (out of) the interval $(x_1,x_2)$ so that they can pass through 
 $A$ ($\grave B$).   \par   We may express this  quantitatively 
  by noting that 
 there
is  now   \cite{Szilard}  a decrease of entropy per molecule 
 after the first step of moving up the pistons.  This  is calculated by 
 taking into 
 account  that now  the  probabilities   to find
 any randomly selected molecule out of (in) the preassigned interval 
 $(x_1,x_2)$   are different from the 
  initial values $w_2$ and $w_1$ before moving up the pistons.  
  Thus, suppose that during
 the first stage of expanding  the initial volume of the cylinder $n_o$ molecules,
 from the total number $n$,  have come out of  the remarked interval and $n_i$
 from outside have entered   so that
 the probability to find now any randomly selected molecule out of it is
 $(w_2+\frac{(n_o-n_i)}{n})$ and that  to find it in  is  
 $(w_1+\frac{(n_i-n_o)}{n})$.  Thus, denoting the  entropies per molecule  
  before and after moving up the pistons by $s_i$ and $s_m$ respectively we have
    \cite{Szilard} 
 \begin{equation} \label{e14} s_i=-k(w_1\ln w_1+w_2\ln w_2), 
 \end{equation}
  \begin{eqnarray}  && s_m=
- k((w_1+\frac{(n_i-n_o)}{n})\ln (w_1+\frac{(n_i-n_o)}{n})+\label{e15} 
\\ &&+(w_2+\frac{(n_o-n_i)}
{n})\ln
 (w_2+\frac{(n_o-n_i)}{n})), \nonumber  
 \end{eqnarray}
 where $k$ is Boltzman's constant. 
 The difference in the entropy per molecule  between the two situations 
 from Eqs
 (\ref{e14})- (\ref{e15}) is \begin{eqnarray} && \delta s=(s_m-s_i)=-(kw_1(\ln
 (w_1+\frac{(n_i-n_o)}{n})-\ln w_1)+kw_2(\ln
 (w_2+\nonumber \\ && +\frac{(n_o-n_i)}{n})- \ln w_2)
 + k\frac{(n_o-n_i)}{n}(\ln (w_2+\frac{(n_o-n_i)}{n})-\label{e16} \\ && - 
   \ln (w_1+\frac{(n_i-n_o)}{n}))) =
    -(kw_1(1+\frac{(n_i-n_o)}{w_1n})
 \ln (1 +
 \frac{(n_i-n_o)}{w_1n})+ \nonumber \\ && +
 kw_2(1+\frac{(n_o-n_i)}{w_2n})\ln (1+\frac{(n_o-n_i)}{w_2n})+ 
  \frac{k(n_o-n_i)}{n}\ln
 (\frac{w_2}{w_1})) \nonumber \end{eqnarray}
  Eliminating $w_2$ through use of 
  the relation $w_1+w_2=1$ one may write the last equation as 
 \begin{eqnarray} &&   \delta s=(s_m-s_i)= 
 -(kw_1(1-\frac{(n_o-n_i)}{nw_1})\ln (1-\frac{(n_o-n_i)}{nw_1})+ \label{e17} \\ 
&& + k(1-w_1)(1+\frac{(n_o-n_i)}{n(1-w_1)})\ln (1+\frac{(n_o-n_i)}{n(1-w_1)})+ 
  \frac{k(n_o-n_i)}{n}\ln
 (\frac{(1-w_1)}{w_1})) \nonumber \end{eqnarray} 
  We note    that the
  probability $w_1$ must be directly  proportional to the length of the remarked interval $x_2-x_1$, 
  so that a small or large value for one indicates a corresponding value for the
  other. Thus, we may  assume a normal distribution   \cite{Spiegel} 
  for $w_1$ in terms of $x$  and write for the density function of $w_1(x)$  
   $f_{w_1}(x)=\frac{\exp(-\frac{(x-\mu)^2}{2\sigma^2})}{\sqrt{2\pi}\sigma}$,  
  where $\mu$ is the mean value of $x$ and $\sigma$ is the standard deviation.
  To further simplify the following calculation we assume a standard normal distribution
    \cite{Spiegel} $z=\frac{(x-\mu)}{\sigma}$ for which  $\mu=0$ and  $\sigma=1$. Thus,  the density function
   $f_{w_1}(x)$ may be written as $ f_{w_1}(z)=
  \frac{\exp(-\frac{(z)^2}{2})}{\sqrt{2\pi}}$ and the probability $w_1(x)$ to
  find any randomly selected molecule in the interval $(-x,x)$, where now this
  interval is symmetrically located around the origin $x=0$, is 
    \cite{Spiegel} 
  \begin{equation} \label{e18} w_1(x)=\int_{-x}^xf_{w_1}(z)dz
  =\frac{1}{\sqrt{2\pi}}\int_{-x}^xdze^{-\frac{z^2}{2}}=
  erf(\frac{x}{\sqrt{2}}) \end{equation} 
 $erf(x)$ is the error function defined as $erf(x)=\frac{2}{\sqrt{\pi}}\int_0^x
 e^{-u^2}du$. Note that $erf(0)=0$, $erf(\infty)=1$, and $erf(-x)=-erf(x)$ so
 that this function is appropriate for a representation of the probability
 $w_1(x)$. Substituting from Eq (\ref{e18}) into Eq (\ref{e17}) we obtain 
\begin{eqnarray} &&   \delta s=(s_m-s_i)= 
 -(k\cdot erf(\frac{x}{\sqrt{2}})
 (1-\frac{(n_o-n_i)}{n\cdot erf(\frac{x}{\sqrt{2}})})
 \ln (1-\frac{(n_o-n_i)}{n\cdot erf(\frac{x}{\sqrt{2}})})+ \nonumber \\ 
 &&+ k(1-erf(\frac{x}{\sqrt{2}}))(1+\frac{(n_o-n_i)}
 {n(1-erf(\frac{x}{\sqrt{2}}))})\ln (1+\frac{(n_o-n_i)}
 {n(1-erf(\frac{x}{\sqrt{2}}))})+ \label{e19}  \\ 
 &&+ \frac{k(n_o-n_i)}{n}\ln
 (\frac{(1- erf(\frac{x}{\sqrt{2}}))}
 {erf(\frac{x}{\sqrt{2}}) })) \nonumber 
  \end{eqnarray}  
  Eq (\ref{e19})  which gives the entropy decrease
  per molecule,  must be multiplied by the number $n$ of molecules in the cylinder
  in order to obtain the total decrease of entropy after moving up the pistons. 
  Figure 4 shows a three-dimensional representation 
   of the entropy $s$  
  per molecule from the
 last equation as function of $\frac{n_i}{n}$ and $\frac{n_o}{n}$  which are
 respectively the fractions of molecules that have entered and come out of the
 interval $(x_1,x_2)$. 
     The probability  $w_1=erf(\frac{x}{\sqrt{2}})$  
must  begin from the  minimum value
 of $\frac{n_0}{n}$  since $w_1$ can 
 not be smaller than $\frac{n_0}{n}$. The ranges of both $\frac{n_i}{n}$ 
 and $\frac{n_o}{n}$ are specified to $0.005 \le \frac{n_i}{n}, \frac{n_o}{n} \le 0.5$
 because in the reversible motion discussed here it is unexpected that more 
 than
 half of the total particles will enter or leave the interval $(x_1,x_2)$. 
 One may realize from the figure that for large  values of 
  $\frac{n_0}{n}$ ($\frac{n_i}{n}$)  and comparatively small values of  
  $\frac{n_i}{n}$ ($\frac{n_o}{n}$) the entopy
  differences tend to $+1$ ($-1$) and when  both $\frac{n_0}{n}$ 
  and $\frac{n_i}{n}$ are large $s$ tends to zero from negative values. \par 
  \begin{figure}[ht]

\begin{minipage}{.50\linewidth}
\centering\epsfig{figure=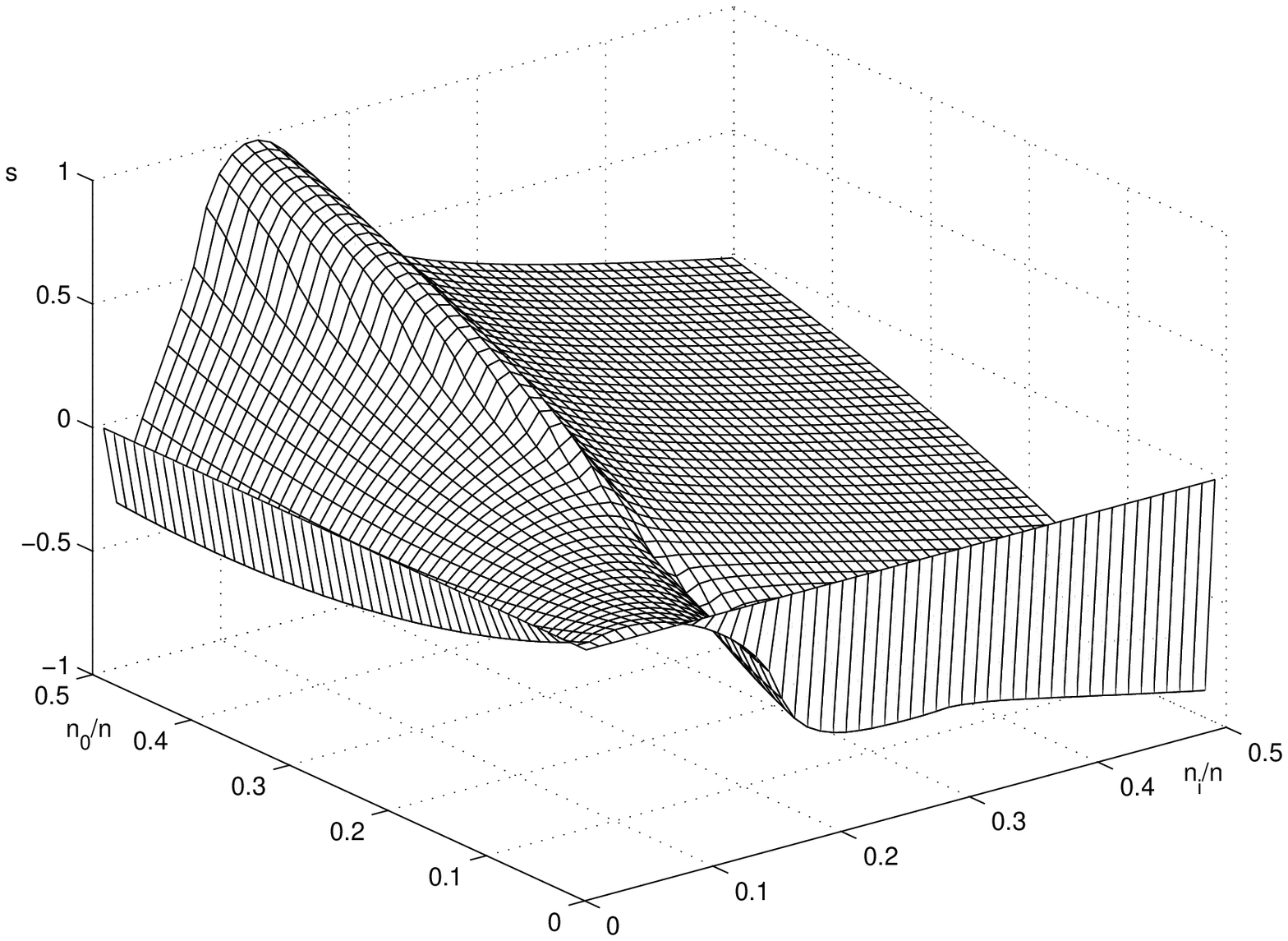,width=\linewidth}

\caption{The figure shows a three-dimensional surface  of the entropy per 
molecule $s$ from Eq (\ref{e19}) 
as function of $\frac{n_o}{n}$ and $\frac{n_i}{n}$.  Both ranges of 
$\frac{n_o}{n}$ and $\frac{n_i}{n}$ are $(0.005, 0.5)$ since it is unexpected
that in a reversible motion more than half of the total molecules will leave or
enter the given interval $(x_1,x_2)$. Note that for large $\frac{n_o}{n}$ 
($\frac{n_i}{n}$) and small $\frac{n_i}{n}$  ($\frac{n_o}{n}$) $s$ tends to $+1$
$(-1)$. }
\end{minipage}
\end{figure}

 As realized from Eq (\ref{e19}) when  $n_o=n_i$,  which means that there is no
 net transfer of molecules out of or into the interval $(x_1,x_2)$, 
 the entropy decrease from Eqs (\ref{e19})  
  is  obviously zero. When, however,  $n_o \ne n_i$ 
  the  molecules that come out of the interval $(x_1,x_2)$ and those
  that have entered it  prevent, as remarked, the reversible return of the
  pistons to their former places.  This problem has been discussed and solved in
  \cite{Szilard} for the single cylinder. Our main interest is to generalize
  from  this four-piston cylinder to a large ensemble of such cylinders and
  calculate, as done for the quantum examples in Sections II-III, the
  correlation among them.   
     \par 
  We  assume that the initial state of all the $N$ identical four-pistons 
  cylinders  is that in which the movable  pistons ${\grave B_j}$, 
 $B_j$ are on the fixed ones   ${\grave A_j}$  and $ A_j$  where $1 \le j \le N$
 (see Figure 3).
 One then simultaneously and reversibly raise up and down in a complete cycle
 all the $2N$ movable pistons ${\grave B_j}$ and $B_j$, $1 \le j \le N$. 
   Thus, if after the moving-up stage  we find,  for some 
 of them,    that no molecule comes out of   
 the interval $(x_1,x_2)$  and  no one from outside has entered it then, as
 remarked, 
 they record no entropy decrease during this stage. Note that if no entropy
 decrease has been detected during the reversible upward motion then one may 
 assume no such
 decrease also in the downward motion. If, on the other hand, one finds    
 $n_o$ molecules come out of the interval $(x_1,x_2)$ and $n_i$ 
 have entered where $n_o \ne n_i$ then, as remarked, a decrease of entropy must
 occurs.   In such case  
  the total decrease of entropy for the $N$
  cylinders after the moving-up stage is   \begin{eqnarray}
 && \delta s_{total}=
 -k\sum_{j=1}^{j=N}n(erf(\frac{x_j}{\sqrt{2}})
 (1-\label{e20} \\ && -\frac{(n_{o_j}-n_{i_j})}{n\cdot erf(\frac{x_j}{\sqrt{2}})})
 \ln (1- 
 \frac{(n_{o_j}-n_{i_j})}{n\cdot erf(\frac{x_j}{\sqrt{2}})})+  
 (1-erf(\frac{x_j}{\sqrt{2}}))(1+\frac{(n_{o_j}-n_{i_j})}
 {n(1-erf(\frac{x_j}{\sqrt{2}}))})\cdot \nonumber \\ && \cdot 
 \ln (1+\frac{(n_{o_j}-n_{i_j})}{n(1-erf(\frac{x_j}{\sqrt{2}}))})
 + \frac{(n_{o_j}-n_{i_j})}{n}\ln
 (\frac{(1-erf(\frac{x_j}{\sqrt{2}}))}
 {erf(\frac{x_j}{\sqrt{2}})})),  \nonumber \end{eqnarray}
  where we use   Eq (\ref{e19}) and  assume   that the total number of 
    molecules $n$  are
 the same for all the ensemble members.  
  We, now,  show that when the  $N$ experiments of reversibly moving the pistons up and down 
 are  related to each other in the  sense 
   that no two of them  share  the same   value  of either 
 $\frac{n_{o_j}}{n}$ or   $\frac{n_{i_j}}{n}$ (or  $x_j$), where  $1 \le j \le N$, 
  then  the    larger  is    $N$ the more probable is to  obtain  
  entropy decrease.  If, on the other hand, they are not related in this manner   
     so that some systems   share  the values of either  $\frac{n_{o_j}}{n}$   
   or   $\frac{n_{i_j}}{n}$ (or $x_j$) 
     then the mentioned probability will be discontinuous, stochastic and 
     much less clear compared to the former case. 
      We first note  that     
   since for all  $x \ge 3$ 
  $erf(x) \approx 1$    we may
  assume a range of $(-3,3)$ from which we take the values for the $N$ 
  preassigned
  intervals $(-x_j,x_j)$ where $1 \le j \le N$.   That is,  we
  subdivide the interval $(-3,3)$ into $N$ different subintervals, 
  where $N$ is   the
  number of  cylinders, so that each  has its unique 
  interval $(-x_j,x_j)$   besides  its specific values of  
  $\frac{n_{o_j}}{n}$ and   
     $\frac{n_{i_j}}{n}$.   Also,    
      each 
  probability $w_{i_j}=erf(\frac{x_j}{\sqrt{2}})$ for any 
  system $O_j,  \ \ (j=1, 2,\ldots, N)$  must begin,  as remarked after Eq
  (\ref{e19}), from the  minimum value 
  of $\frac{n_{o_j}}{n}$ and we also assume 
  (see the discussion after Eq (\ref{e19})) 
   that the $2N$ different values of   $\frac{n_{i_j}}{n}$ and 
   $\frac{n_{o_j}}{n}$, $1 \le j \le N$  are from 
    the range $0.005 \le \frac{n_{o_j}}{n}, \frac{n_{i_j}}{n} \le 0.5$.  
   We assign to each experiment that results in  entropy decrease,  after
   moving-up the pistons,  the value of $+1$ 
    and 0 otherwise.  Thus,   
   assuming that the movable pistons in the $N$ cylinders are moved up 
     we calculate the quantity \begin{equation} \label{e21} 
   g(N)=\frac{1}{N}\sum_{i=1}^{i=N}g_i(N), \end{equation}  
   where $g_i(N)=1$ for an entropy decrease result 
 and $g_i(N)=0$ otherwise.  That is, the function $g(N)$ is directly proportional 
 to the number of experiments which result in entropy decrease and inversely
 proportional to those with a different result (for which $\delta s \ge 0$). 
  Figure 5  shows  $g(N)$    as a  function of  $N$, in the range $400 \le N \le 3500$, 
    and we see that $g(N)$ grows as the number $N$ of related  cylinders 
  increases where this relationship is effected, as remarked,  by  preparing the $N$ 
  experiments so that any one of them have its unique  $\frac{n_{o_j}}{n}$, 
 $\frac{n_{i_j}}{n}$   and $(-x_j,x_j)$  where $ 1 \le j \le N$.    
  That is, the  larger is the number of related 
  experiments the more frequent is the result of entropy decrease.   
   If, on the other hand, this kind of
  relationship  is absent as when   
   assigning randomly to any system  $O_j \ \ (j=1, 2, \ldots, N)$ an
  interval $(-x_j,x_j)$   (from $(-3,3)$) and also $\frac{n_{o_j}}{n}$,   
  $\frac{n_{i_j}}{n}$ (from $(0.005, 0.5)$)  we obtain a stochastic result for 
  $g(N)$    that
  implies no clear-cut consistent value.  
  This is clearly seen 
    in  the sawtooth
  form of the curve of Figure
  6 which is drawn under exactly the same conditions as those of  Figure 5
   except that the values of $(-x_j,x_j)$, $\frac{n_{o_j}}{n}$ and  
  $\frac{n_{i_j}}{n}$ are randomly chosen.  \par 
  
  \begin{figure}[ht]
\begin{minipage}{.48\linewidth}
\centering\epsfig{figure=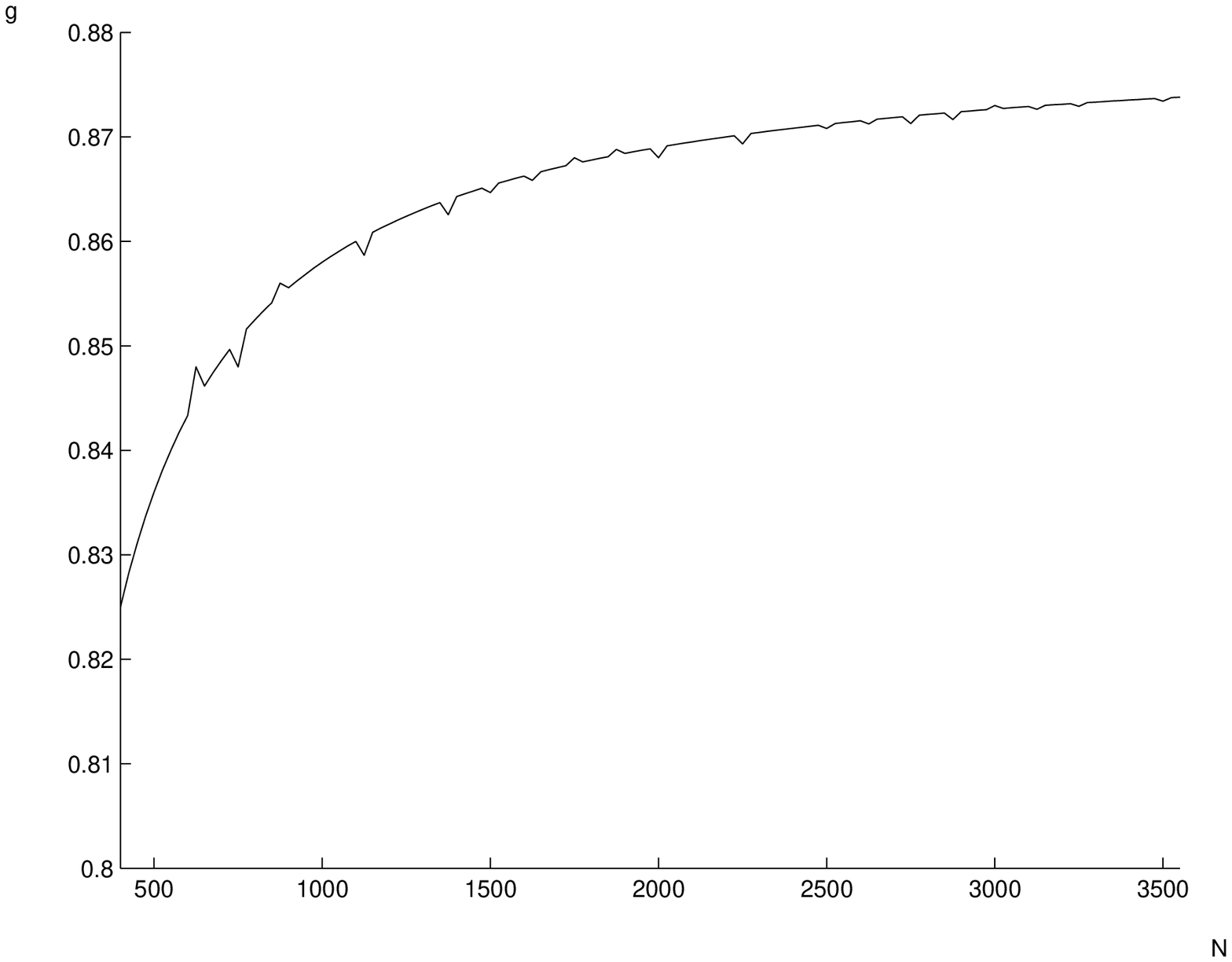,width=\linewidth}

\caption{The curve shows the form of  $g(N)$ from  
 Eq  (\ref{e21}) as a function of $N$ 
 after   performing the $N$  
experiments  of lifting up the pistons where 
 $400 \le N \le 3500$.     Note that no two of the
$N$  experiments are  identical and that each is deliberately performed 
 for different  values of 
 $(-x_j,x_j)$, $\frac{n_{o_j}}{n}$ and $\frac{n_{i_j}}{n}$ where 
$x_j=6\cdot \frac{n_{o_j}}{n}$.  We see  
that as  $N$ grows the number of experiments that end in an  entropy
decrease increases.}
\end{minipage}  \hfil
\begin{minipage}{.48\linewidth}
\centering\epsfig{figure=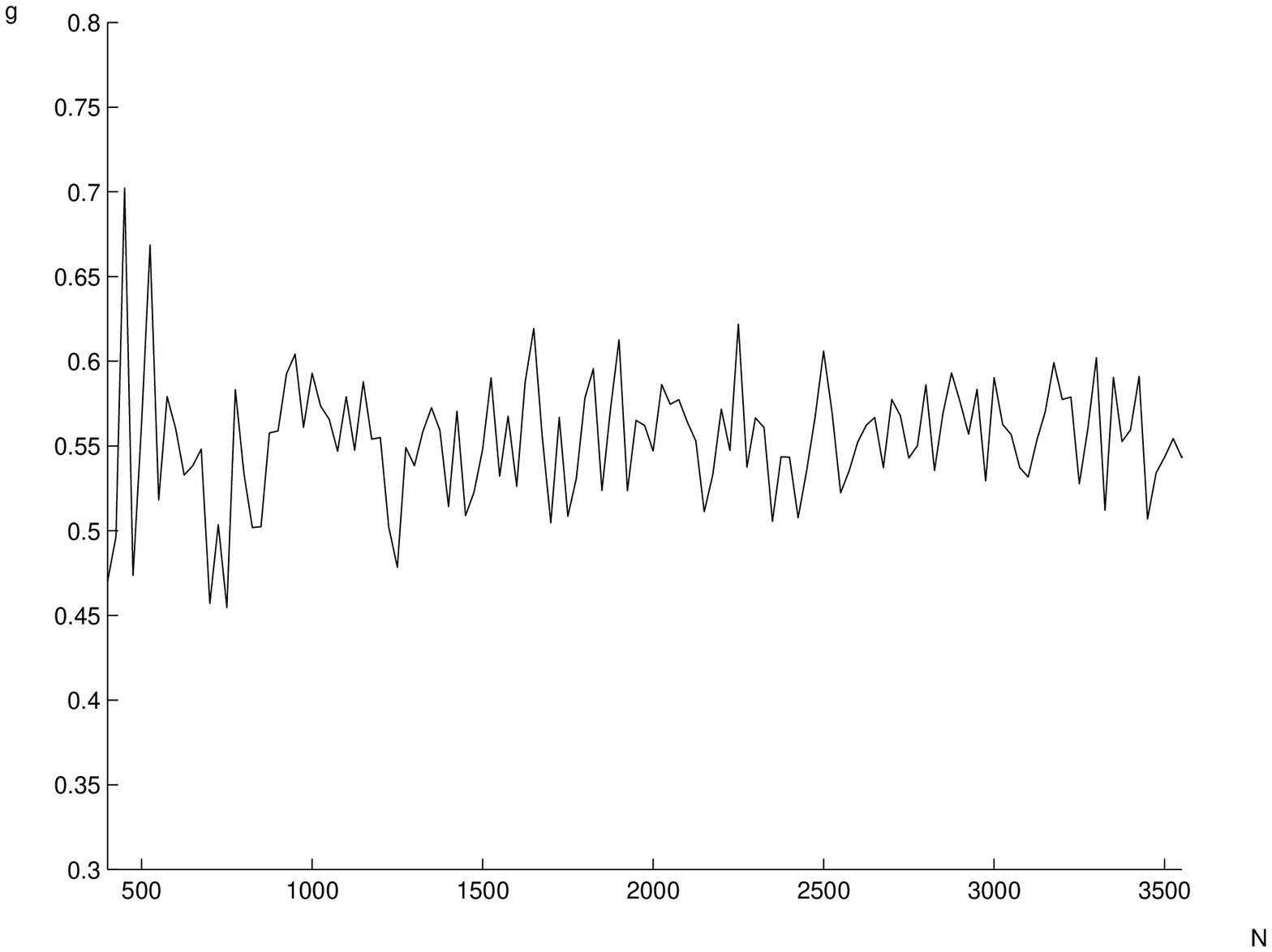,width=\linewidth}

\caption{The stochastic graph,   which shows   
 $g(N)$ from   
Eq (\ref{e21}) as a function of $N$, is drawn for exactly the same
conditions as those of Figure 5  except that the values of   
$\frac{n_{o_j}}{n}$   and $\frac{n_{i_j}}{n}$ are randomly chosen.    
 Note  that in contrast to  Figure 5 some of these experiments  may be 
 identical
due to the random conditions under which they are performed. Thus, the results
do not  show any clear-cut consistent value for the entropy differences. }
\end{minipage}
\end{figure}
   We note that the same results may be obtained by using other methods 
  and terminology.
  Thus, 
   it is shown  \cite{Gisin2} that the ``localization''  (in the
 sense of smaller dispersion) for the state $|\phi\!>$ is greater the  smaller 
  is
 the entropy which results when  the rate of ``effective interaction
 with the environment''  \cite{Gisin2} increases.  Localization is another 
 name    for 
 what we call here
 ``realizing or preserving a specific state''   and the interaction with the 
 environment is equivalent to
 performing  experiment   \cite{Harris,Davies,Zeh}, so that as the rate of performing
 experiment grows the more realized and  
  localized is the state one  begins with or the path of states along which one
 proceeds. 
 
 \bigskip \bigskip 
  
\protect \section*{Concluding Remarks}     
  We have studied the influence of obsevation, and especially the large number
  of them, upon the obtained results. 
  This has been shown  for
 both  quantum and classical systems. For the quantum part in Sections 2-3 we
 have made use of the Feynman path integral  \cite{Feynman} and 
 the Everett's relative
 state  \cite{Everett,Graham} methods. For the classical part in Section 4 we
 use entropy considerations \cite{Reif} for discussing the four-piston cylinder
 \cite{Szilard}. Using these  analytical methods we show that for producing the
 obtained results all the involved systems and experiments should be related to
 each other in some kind of relationship which assumes different, and even
 contradictory, forms for different situations. Thus, for the static Zeno effect
 the relationship between the systems is  their being initially prepared in the
 same initial state and for the dynamic Zeno and the classical cylinder this
 relationship is effected by initially preparing the systems in different
 states.   \par
 This is, especially, emphasized 
 in   a clearer way using 
  entropy considerations in Section 4. 
  The important factor that entails the collective  entropy decrease 
     is,  as remarked,  
   when  all the memebers of the ensemble are related to each other as described
   in Section 4  (see Figure 5). Unrelated ensemble of observers, 
 no matter how
 large it is, does not obtain the same required entropy decrease as 
 seen clearly in Figure 6. \par 
   
 \noindent \protect \section*{\bf Acknowledgement }
  \bigskip  \noindent  I wish to thank L. P. Horwitz for discussions on this 
subject.

\bigskip \bibliographystyle{plain}

\end{document}